\documentclass[prl,a4paper,superscriptaddress,twocolumn,showpacs,amsmath,amssymb,floatfix,amstex]{revtex4}

\usepackage{psfrag}

\usepackage[dvips]{graphicx}
\usepackage{bm,pstricks,longtable}

\begin{document}

\title{Dynamical Thermalization of Disordered Nonlinear Lattices}
\author{Mario Mulansky} 
\affiliation{\mbox{Department of Physics and Astronomy, Potsdam University, 
  Karl-Liebknecht-Str 24, D-14476, Potsdam-Golm, Germany}}
\author{Karsten Ahnert} 
\affiliation{\mbox{Department of Physics and Astronomy, Potsdam University, 
  Karl-Liebknecht-Str 24, D-14476, Potsdam-Golm, Germany}}
\author{Arkady Pikovsky} 
\affiliation{\mbox{Department of Physics and Astronomy, Potsdam University, 
  Karl-Liebknecht-Str 24, D-14476, Potsdam-Golm, Germany}}
\affiliation{\mbox{Universit\'e de Toulouse, UPS,
Laboratoire de Physique Th\'eorique (IRSAMC), F-31062 Toulouse, France}}
\author{Dima L. Shepelyansky}
\affiliation{\mbox{Universit\'e de Toulouse, UPS,
Laboratoire de Physique Th\'eorique (IRSAMC), F-31062 Toulouse, France}}
\affiliation{\mbox{CNRS, LPT (IRSAMC), F-31062 Toulouse, France}}

\date{\today}

\begin{abstract}
\noindent We study numerically how the energy spreads over a finite 
disordered nonlinear one-dimensional lattice, where all linear modes
are exponentially localized by disorder. We establish emergence of
dynamical thermalization, characterized as an ergodic chaotic
dynamical state with a Gibbs distribution over the modes. 
Our results show that the fraction of thermalizing modes is finite and 
grows with the nonlinearity strength.
  
\end{abstract}


\maketitle

\noindent The studies of ergodicity and dynamical 
thermalization in {\em regular}
nonlinear lattices have a long history initiated 
by the Fermi-Pasta-Ulam problem \cite{fpu} 
but they are still far from being complete (see, e.g., \cite{lepri}
for thermal transport in nonlinear chains and
~\cite{cassidy09}
for thermalization in a Bose-Hubbard model). In this letter, we study how the
dynamical thermalization appears in nonlinear {\em disordered} chains
where all linear modes are exponentially localized.
Such modes appear due to
the Anderson localization, introduced  in the context of 
electron transport in disordered solids~\cite{Anderson-58,Lee-Ramakrishnan,Kramer-MacKinnon-93} and describing various physical 
situations like wave propagation 
in a random medium~\cite{Sheng-06}, 
localization of a Bose-Einstein condensate~\cite{aspect2008}  and 
quantum chaos~\cite{Fishman-Grempel-Prange-87}.

Effects of nonlinearity on localization properties have
attracted large interest recently. Indeed, nonlinearity naturally appears for localization of a
Bose-Einstein condensate, as its evolution is described by the nonlinear 
Gross-Pitaevskii equation~\cite{Dalfovo-99}.
An interplay of disorder, localization, and nonlinearity
is also important
in other physical systems like
wave propagation in nonlinear disordered
media~\cite{Skipetrov-00,Lahini-08}
and
chains of nonlinear oscillators
with randomly distributed frequencies~\cite{Dhar-Lebowitz-08}.

The main question here is whether the localization is destroyed by nonlinearity.
It has been addressed recently using two physical setups. 
In refs.~\cite{Molina-98,Pikovsky-Shepelyansky-08}
 it was demonstrated
that an initially concentrated wavepacket spreads apparently indefinitely, although subdiffusively, 
in a disordered nonlinear lattice. 
For a transmission through a nonlinear disordered 
layer~\cite{Paul-05,Tietsche-Pikovsky-08},  chaotic destruction
of localization  leads to a drastically enhanced transparency. 

Here we study the thermalization properties of the dynamics of a nonlinear
disordered lattice -- discrete Anderson nonlinear Schr\"odinger equation (DANSE). 
We describe in details the features of the  time-evolution
of an initially localized excitation towards a statistical equilibrium in a finite lattice (we stress that this evolution is purely deterministic -- and the relaxation to equilibrium is due to determinsitc chaos.).
Below we argue that a statistically stationary state is characterized by the Gibbs energy equipartition across the linear eigenmodes (Eq.~(\ref{eq6})) and call a relaxation to such an equilibrium state thermalization. Because thermalization is due to deterministic chaos, its rate heavily dependes on the statistical properties of the chaos. As is typical for nonlinear Hamiltonian systems, depending on initial conditions one can obtain solutions belonging to a ``chaotic sea'' or to ``regular islands''. Moreover, one can expect the former to thermalize while the latter do not lead to thermalization.
We numerically find non-thermalizing modes and characterize 
their dependence on the nonlinearity and the lattice length. 
We stress here that our analysis  heavily relies on numerical simulations as analytic methods appear to be hardly applicable for disordered nonlinear systems. In numerics, a difference between thermalizing and non-thermalizing states (as well as between chaotic and non-chaotic states) is limited by the maximal integration time: it might happen that the states which do not thermalize up to some time will thermalize in the future. There is no way to overcome this  limitation in a simple way, because of a possibility for such slow processes like Arnold diffusion, characteristic time of which lies far beyond any computationally accessibility. Nevertheless, peforming an analysis based on large but finite time scales, we can, on one hand, 
make predictions for experiments, and on the other hand, obtain a ``coarse-grained'' description of the dynamics. Accordingly, the results below should be understood as valid for available integration times, without a straightforward extrapolation for asymtotically large times.

We describe a nonlinear disordered medium by the DANSE model:
\begin{equation}
i \frac{\partial {\psi}_{n}}{\partial {t}}
=E_{n}{\psi}_{n}
+{\beta}{\mid{\psi_{n}}\mid}^2 \psi_{n}
 +{\psi_{n+1}}+ {\psi_{n-1}}\;,
\label{eq1}
\end{equation}
where $\beta$ characterizes nonlinearity, and the on-site
energies $E_n$ (or frequencies)
are independent random variables distributed uniformly
in the range $-W/2 < E_n < W/2$ 
(they are  shifted in such a way that $E=0$
corresponds to the central energy of the band).
We consider a finite lattice $1\leq n\leq N$ with periodic boundary conditions. Then
DANSE is a classical dynamical system with the Hamilton function
\begin{equation}
H=\sum_{n}E_n|\psi_n|^2+\psi_{n-1}\psi_n^*+\psi_{n-1}^*\psi_n+\frac{\beta}{2}|\psi_n|^4\;.
\label{eq2}
\end{equation}
It describes recent experiments with 
nonlinear photonic lattices (cf. Eq.~(1) in \cite{Lahini-08}), where
 one follows, along a transversally disordered,  finite nonlinear crystal, the
 evolution of a single-site or a single-mode initial state. This corresponds
 to the setup of our thermalization problem. Thus, the properties below can be
 observed experimentally as ``thermalization of photons'', provided the
 crystal is long enough. In the context of many-particle quantum systems,
 Eq.~(\ref{eq2}) is used as an effective mean-field Hamiltonian of interacting bosons.

For $\beta=0$ all eigenstates are exponentially localized
with the localization length $l \approx 96 W^{-2} $ (for weak disorder)
at the center of the energy band \cite{Kramer-MacKinnon-93}. 
Below we mainly focus on the case of moderate disorder $W=4$, 
for which $l\sim 6$ at the center of the band and $l\approx 2.5$ at $E = \pm 2$. 
For each particular realization of disorder
a set of eigenergies 
$\epsilon_m$ and of corresponding eigenmodes $\varphi_{nm}$ can be found.
In this eigenmode representation $\psi_n=\sum C_m\varphi_{nm}$ 
the Hamiltonian reads
\begin{equation}
H=\sum_{m} \epsilon_m |C_m|^2+\beta\sum_{knji}V_{knji}C_{k}C_{n}C^*_{j}C^*_{i},
\label{eq3}
\end{equation}
with $\sum_m |C_m|^2=1$ and 
$V_{{m}{m'}{m_1}{m_1'}} \sim l^{-3/2}$ are the transition
matrix elements \cite{dls1993}.
This representation is mostly suitable to characterize the spreading of the field over the lattice,
since in this basis the transitions take place only due to nonlinearity.
Also, the nonlinear correction
to the energy is small ($\sim \beta/l$)  for one excited mode.

\begin{figure}[tbh]
   \centering
   \includegraphics[width=0.42\textwidth,angle=0]{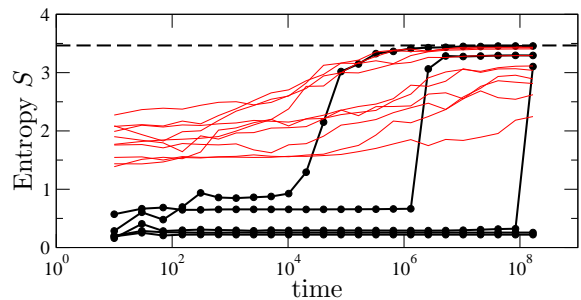}
\vglue -0.3cm
\caption{(color online) Time evolution of entropy $S$ (\ref{eq4}) 
in DANSE (\ref{eq1}) with $N=32$ and $\beta=1$, 
for a particular realization of disorder and different initial states: 
bold black curves with markers -- single-mode initial states (B)
with energies $E=-0.34, 0.76, -0.29, 3.36, -0.5$ (curves from top to bottom at $t=10^8$,
two bottom cases are very close), 
solid red/gray curves -- single-site initial states (A, ten randomly chosen states). 
The dashed line shows the level $S=\ln 32$. The time
averaging has been performed over doubling time intervals (between successive markers).
\label{fig1}
}
\end{figure}

To study the dynamical thermalization in a lattice, we performed direct numerical simulation 
of DANSE~(\ref{eq1}), using mainly two methods: the unitary Crank-Nicholson operator splitting 
scheme with step $\Delta t=0.1$ as described 
in \cite{Pikovsky-Shepelyansky-08}, 
and an 8-order Runge-Kutta integration with step $\Delta t=0.02$; in both cases the total 
energy and the normalization have been preserved with high accuracy and both integration schemes gave similar results,  for all lattice lengthes $N$ used. Such a restriction of the accuracy check to the conserved quantities is suitable for chaotic systems.
A comparison with other numerical methods for DANSE~\cite{jpa} goes beyond the scope of this letter and will be performed in a longer publication. 
We started with two types of
localized initial states: (A) one site seeded, i.e. $|\psi_n(0)|^2=\delta_{n,j}$ and (B) one
mode initially excited, i.e. $|C_m(0)|^2=\delta_{m,k}$. For different realizations of disorder,
we seeded different possible sites/modes and followed the evolution of the field solving (\ref{eq1})
up to times (in selected runs) $\sim 10^8$. The level of spreading is characterized by the entropy
of the mode distribution
\begin{equation}
S=-\sum \rho_m\ln\rho_m\;,\qquad \rho_m=\overline{|C_m|^2}\;,
\label{eq4}
\end{equation}
where overline means time averaging. For one excited mode $S=0$ while $S=\ln N$ for 
a uniform distribution over all modes in a lattice of length $N$.
To give an impression of a time evolution of the thermalization process we show
in Fig.~\ref{fig1} several representative
time dependencies of the entropy (\ref{eq4}). 
One can see that for the setup (B)
some modes remain localized during the complete integration time 
(cf.~\cite{Kopidakis-Aubri-00}), 
while other after some transient time evolve to a state with
large entropy. For setup (A), the entropy grows in all cases with a tendency to saturation --  some states seem to saturate at 
about $S \approx \ln N$, while others remain at values definitely smaller than $\ln N$ up to the maximal integration time. Especially the results from (B) indicate a strong energy dependence of the spreading behavior, which is studied in this work. In our discussion below we focus 
therefore on the setup (B) as the mostly nontrivial one.

To derive an approximate expression for the statistically stationary distribution $\rho_m$,
we mention that it should satisfy
$\sum\rho_m=1$ and  $E=\sum \rho_m\epsilon_m$,
where, in view of discussion above, we have neglected the nonlinear contribution to the energy. 
Then the condition
of maximal entropy (\ref{eq4}) leads, 
after a standard calculation, to a Gibbs distribution:
\begin{equation}
\rho_m=Z^{-1}\exp(- \epsilon_m/T),\qquad Z=\sum_m \exp(- \epsilon_m/T).
\label{eq6}
\end{equation}
Here $T$ is an effective ``temperature'' 
of the system: it has no meaning as a physical temperature, but serves as a parameter characterizing the distribution; it is a function of the total energy $E$ of the state and of the realization of disorder.
The entropy and the energy are related to each other via usual expressions, e.g.~\cite{landau}:
\begin{equation}
E= T^2 \partial \ln Z/\partial T\;, \qquad S= E/T+\ln Z\;.
\label{eq7}
\end{equation}
This value of entropy yields the maximal possible equipartition for the given initial
energy, and the values of Fig.~\ref{fig1} obtained via a numerical simulation of the
disordered nonlinear lattice should be compared with these values from the Gibbs distribution.
Because we have anyhow neglected the effects of nonlinearity in the theoretical value of the
entropy, we adopt other simplifications: approximate the density of states of the disordered
system as a constant in an interval $-\Delta<\epsilon<\Delta$ 
and consider the energy eigenvalues $\epsilon_m$ in
a particular realization of disorder as independent 
random variables distributed according to this density.
Furthermore, we assume the variations of the partition sum to be small and use
$\langle \ln Z\rangle\approx \ln \langle Z\rangle$, where brackets denote averaging over 
disorder realizations. In this  simplest approximation we obtain
\begin{equation}
\langle \ln Z\rangle\approx \ln N+\ln\sinh(\Delta/T)-\ln(\Delta/T) .
\label{eq8}
\end{equation}
At $W=4$ we have $\Delta\approx 3$ (see Figs.~\ref{fig3},\ref{fig5} below) and this theory
gives the dependence $S(E)$ within a few percent 
accuracy compared to
$S$ averaged over disorder within
Gibbs computations with exact numerical values $\epsilon_m$.
 This justifies, for the parameters used, the approximation above.
We note that $T=+0, \pm\infty, -0$ correspond to
 $E=-\Delta, 0, +\Delta$ respectively (as in the standard two-level problem, see related discussion in \cite{landau}).

\begin{figure}[tbh]
   \centering
\includegraphics[width=0.42\textwidth,angle=0]{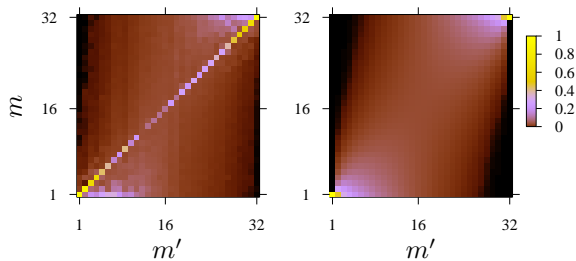}
\vglue -0.3cm
\caption{(color online) Left:
time and disorder averaged probability  
$\langle \overline{\rho_m(m')}\rangle$ in mode $m$ 
for initial state in mode $m'$. 
Right: 
theoretical values according to the Gibbs distribution (\ref{eq6}).
Here $N=32, \beta=1, N_d=15$.
\label{fig2}
}
\end{figure}

\begin{figure}[tbh]
   \centering
   \includegraphics[width=0.42\textwidth,angle=0]{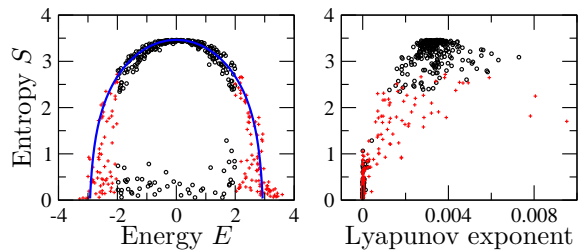}
\vglue -0.3cm
\caption{(color online) Left panel: Final entropies (\ref{eq4}) after an
  evolution during time interval $10^7$, averaged over a time interval of
  $10^6$. The states evolving from initial modes in the middle of the band
  (see text) are marked with black circles, while those at the edges of the
  band are marked by the red (gray) pluses.  The curve shows the approximate
  theory (\ref{eq8}).  Right panel: Lyapunov exponents `` (averaged over a
  time interval $10^6$) vs. entropy for the same sets. Here $N=32$, $\beta=1,
  N_d=7$.
\label{fig3}
}
\end{figure}

We compare in Fig.~\ref{fig2} the Gibbs distribution (\ref{eq6}) 
with the results of direct numerical simulations of DANSE
using $N_d$ disorder realizations. 
Here we present the values $\langle\overline{\rho_m}\rangle$ averaged over time 
and over different realization of disorder, in dependence of the number 
of the initially seeded mode $m'$. The modes have been ordered according to their energy,
so that $m=1$ corresponds to the maximal energy. One can see a good correspondence 
between the numerics and the simple theory (\ref{eq6}) in the sense that states at the band edges remain localized, while states in the center generally spread. However, there is one
crucial discrepancy: the peaks on the diagonal $m=m'$ 
indicate that there are cases when there is 
no thermalization within our simulation time and the energy remains in the initially seeded mode.

To characterize thermalized and non-thermalized cases quantitatively,
we compare in Fig.~\ref{fig3} numerical values for $S(E)$ according to Eq.~(\ref{eq4}) with
the theoretical Gibbs computation given by Eqs.(\ref{eq6},\ref{eq7},\ref{eq8}).
Clearly, the Gibbs theory gives a satisfactory global description of numerical data. The nonthermalized modes in this presentation are those at the bottom of the graph; these states are absent for the setup (A) where initial sites are seeded. (Again, as discussed above, ``nonthermalized'' should be understood as ``nonthermilized whithin the integration time'').

It appears appropriate to discuss the dynamics of the modes in the middle of 
the energy band ($|\epsilon_m|<2$) and at the edges ($|\epsilon_m|>2$)  separately.  For the modes in the middle of the  band,
the maximal entropy according to (\ref{eq7}) is close to $\ln N$, and 
one clearly distinguishes the thermalized modes and those that 
remain localized, as those reaching the maximal entropy and those remaining at the level 
$S\lesssim 1$, correspondingly. 
Thermalization is associated with 
the chaotic dynamics of the DANSE lattice. To demonstrate this, 
we calculated the largest Lyapunov exponents $\lambda$
shown in Fig.~\ref{fig3} (right panel).
All modes with $S< 1$, 
i.e. those that do not thermalize, have nearly vanishing $\lambda$, 
while for the thermalized states ($S > 2$) the positive values of 
$\lambda$ clearly indicate chaos.

The above distinction between thermalized and non-thermalized states is less
evident for modes at the band edges (shown by red pluses in Fig.~\ref{fig3}). 
Here already the theoretical value of entropy given by Eqs.~(\ref{eq6},\ref{eq7},\ref{eq8}) 
is rather small. Hence, the spreading can go over a few ``available'' modes only. 
Nevertheless, also here one can see from Fig.~\ref{fig3} a clear correlation between 
the entropy and the Lyapunov exponent. Moreover, in several cases the Lyapunov exponent 
at the edge of the spectrum is definitely larger than in the middle.  
This happens because the energy spreads over a small number of modes, 
hence the effective nonlinearity is larger due to larger amplitueds of each mode, and therefore chaos is stronger.

Above, we did not account for a spatial organization of the mode structure. 
The latter is less important for the modes in the middle of the band, where one can 
always expect to find neighbors with a close energy. Contrary to this, 
for the energies at the edges the issue of spatial distance becomes essential. 
Indeed, since here the thermalization is possible only over a few modes, 
it is important whether these modes are spatially separated or not. For linear eigenmodes $m$ and $m'$
the natural measure
of this separation 
is the coupling matrix element $V_{m'm'm'm}$ according to (\ref{eq3}). 
It is exponentially small 
for spatially separated modes due of their localization. One can expect 
that a mode at the edge of the spectrum will be thermalized only if the coupling $V$ 
to other few modes in the lattice with a close energy is  large,
what is a rather rare event. 

\begin{figure}[tbh]
   \centering
   \includegraphics[width=0.42\textwidth,angle=0]{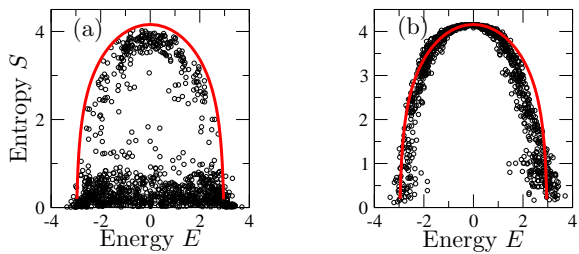}
\vglue -0.3cm
\caption{(color online) Dependence of entropy $S$ on 
energy $E$ as in Fig.~\ref{fig2} but for $N=64, N_d=18$, and 
two values of nonlinearity: (a) $\beta=0.5$, (b) $\beta=2$. 
Averaging have been performed over the time interval $10^6$ after 
an initial evolution during time $10^6$; for small $\beta$ 
still longer times are needed to reach thermalized state with 
maximal $S$ at given $E$.
\label{fig5}
}
\end{figure}

\begin{figure}[tbh]
   \centering
    \psfrag{xlabel}[c][c]{$\beta$}
    \psfrag{ylabel2}[c][c]{$f_{th},f_{b}$}
    \psfrag{ylabel1}[c][c]{$\epsilon_m$}
\psfrag{a}{(a)}
\psfrag{b}{(b)}
   \includegraphics[width=0.42\textwidth,angle=0]{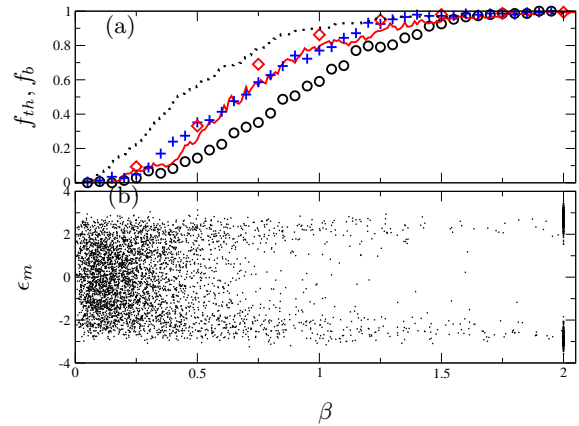}
\vglue -0.3cm
\caption{(color online)(a) Fraction of thermalized (after time $10^6$) modes 
$f_{th}$ from the middle of the band as a function of nonlinearity  $\beta$ for
$N=16$ (circles), $32$ (bold line), and $64$ (pluses). Diamonds show data  for $t=10^7$ and $N=32$. The dotted line shows the fraction of the bifurcated breathers $f_b$ according to panel (b). Panel (b): the bifurcation values  $\beta_c$ for different modes vs their linear energies $\epsilon_m$ for $N=32$. 
To all modes with $\beta_c>2$ we have attributed $\beta_c=2$, this set looks like
two vertical ``lines'' at $\beta=2$ on panel (b).
\label{fig6}
}
\end{figure}

Finally, we discuss how the thermalization properties depend on the nonlinearity constant $\beta$. 
In Fig.~\ref{fig5} we show the dependence $S(E)$ 
for different nonlinearities $\beta$.  
For $\beta=0.5$ a large portion of the initial states remains non-thermalized, 
while for $\beta=2$ all states are thermalized
(at least in the sense that their entropy is close to the maximal possible one, 
as discussed above this is a good criterion in the middle of the band).
To determine how the fraction of thermalized
states depends on nonlinearity $\beta$ we use the following procedure.
For the initial modes in 
the middle of the band (i.e. for $|E|<2$) we classified those that reach 
more than the half of the maximal entropy
(i.e. the level $\ln(N)/2$) as thermalized, and those 
that remain below this level as non-thermalized. The fraction $f_{th}$ of the thermalized modes, 
shown in Fig.~\ref{fig6}, monotonously increases with $\beta$. 
At fixed $\beta$ the numerical data indicate saturation of $f_{th}$ at large $N$,
but more detailed checks at larger sizes and longer times are required. 
For example, recent results on self-induced transparency 
of a disordered nonlinear layer~\cite{Tietsche-Pikovsky-08}
show decrease of critical $\beta$ with lattice size for $N \leq 32$.

The properties of thermalization described above can be incorporated in a general framework of nonlinear dynamics as follows. One can expect, based on general KAM arguments, that for small nonlinearity regular, non-ergodic regimes predominate, while for large values of $\beta$ stable solutions are destroyed and a chaotic ergodic state establishes in the lattice. While it is hard to characterize this transition via a general analysis of the dynamics in a high-dimensional phase space, it is possible to follow the evolution, as nonlinearity increases, of special resonant modes that stem from linear ones. Looking for solutions of (\ref{eq1}) in the form $\psi_n(t)=\phi_ne^{-i\epsilon t}$, we arrive at a nonlinear eigenvalue problem $\epsilon \phi_n=E_n\phi_n+\beta \phi_n^3+\phi_{n-1}+\phi_{n+1}$ which, of course, at $\beta=0$ yields linear frequencies and modes. Starting from these modes, we followed these solutions to larger nonlinearities using a numerical continuation, and in this way obtained nonlinear resonant modes -- ``breathers'' (cf.~\cite{Kopidakis-Aubri-00,Iv}).
Worthnoting, these modes change in the regions where the field is large, while
in the tails they follow linear solutions, in accordance with~\cite{Iomin-Fishman-07}.
 Moreover, we performed numerical stability analysis of these breathers and found that they bifurcate at some critical value of nonlinearity $\beta_c$. The values of $\beta_c$ for an ensemble of realizations of random potentials are shown in Fig.~\ref{fig6}b. Additionally, we show in Fig.~\ref{fig6}a a cumulative distribution of $\beta_c$ for the same range of eigenenergies $|\epsilon_n|<2$ that is used for the other curves plotted. 
First of all, note the similar global behavior of $f_\text{th}$ and $f_\text{b}$ which makes us believe that the bifurcations of stable resonant modes is indeed the mechanism of the $\beta$-dependence of thermalization.
However, the curves do not coincide because $\beta_c$ is defined as the value of the first bifurcation, which may not immediately lead to chaos but may be the first one in a series of transitions to more irregularity.
Strictly speaking, $f_\text{b}$ should be an upper bound for $f_\text{th}$, which is seen in Fig.~\ref{fig6}a.
The increase of $f_\text{th}$ from $t=10^6$ to $10^7$ shows that it hasn't saturated yet, but the saturation curve must lie below $f_\text{b}$. 
Remarkably, we have found that the breathers at the edges of the band, i.e. for $|\epsilon_n|>2$, are extremely stable: most of them remain stable up to large values of $\beta\approx 5$. This corresponds to the numerical observation of the strong suppression of the thermalization for these modes.
We emphasize here that because of the nonlinearity of the system the superposition principle does not hold. This means that to observe a stable breather mode one has to prepare initial conditions mostly close to this solution -- what is achieved here by choosing the initial conditions as a pure linear eigenmode (case B above). 
When one initially seeds one site, as in case A (or uses other initial conditions not close to a breather), then this initial condition does not produce a breather because the latter typically does not survive nonlinear interaction with other components of the solution.  
If, for example, one starts with an excitation of two modes which are both stable at some value of $\beta$, one might still see fast thermalization, because a superposition of two breathers is not a breather.

Our main conclusion  is that the maximally thermalized state in a disordered nonlinear lattice (\ref{eq1}), that 
emerges as a result of chaotic dynamics, is described by 
the Gibbs distribution over the linear modes, 
with some effective {\it temperature} depending on the
initial excitation. 
Not all modes lead to thermalization, some fraction of them remains 
localized, but this fraction decreases with nonlinearity.
We found that this can be explained by the disappearance (via bifurcations) as the nonlinearity increases, of stable resonant modes -- breathers -- stemming from linear eigenstates.
Further studies are still required to establish the properties of 
 this thermalization in dependence  on the nonlinearity
strength, disorder and lattice size.

\vglue -1.5em


\end{document}